\documentclass[useAMS, iop]{emulateapj}
\usepackage{graphicx, hyperref, color}
\usepackage{natbib}
\usepackage{amsmath,amssymb}
\usepackage{aas_macros}

\newcommand{\el}[2]{\ensuremath{^{#1}\mathrm{#2}}}

\newcommand{\Teff}{$T_{\rm eff}$}
\long\def\symbolfootnote[#1]#2{\begingroup%
\def\thefootnote{\fnsymbol{footnote}}\footnote[#1]{#2}\endgroup}

\begin{document}

\shorttitle{LITHIUM IN GC GIANTS}
\shortauthors{D'ORAZI ET AL.}

\title{Lithium abundances in globular cluster giants: NGC 6218 (M12) and NGC 5904 (M5)\footnotemark[1]}
\footnotetext[1]{Based on observations taken with ESO telescopes under program 087.D-0276(A)}

\author{ Valentina D'Orazi\altaffilmark{2,3}, George C. Angelou\altaffilmark{3}, Raffaele G. Gratton\altaffilmark{4}, John C. Lattanzio\altaffilmark{3}, Angela Bragaglia\altaffilmark{5}, Eugenio Carretta\altaffilmark{5}, Sara Lucatello\altaffilmark{4} and Yazan Momany\altaffilmark{4,6}}

\altaffiltext{2}{Department of Physics and Astronomy, Macquarie University, Sydney, NSW 2109, Australia.} 
\altaffiltext{3}{Monash Centre for Astrophysics,  School of
Mathematical Sciences, Monash University,  Melbourne,  VIC 3800,  Australia.}
\altaffiltext{4}{INAF - Osservatorio Astronomico di Padova, Vicolo dell'Osservatorio 5, 35122, Padova, Italy}
\altaffiltext{5}{INAF - Osservatorio Astronomico di Bologna, via Ranzani 1, 40127, Bologna, Italy }
\altaffiltext{6}{European Southern Observatory, Alonso de Cordova 3107, Vitacura, Santiago, Chile}

\vspace{1em}
\email{valentina.dorazi@mq.edu.au}

\begin{abstract}
Convergent lines of evidence suggest that globular clusters host multiple stellar populations. 
It appears that  they experience at least two episodes of star formation whereby a fraction of first-generation stars contribute astrated ejecta to form the second generation(s). 
To identify the polluting progenitors  we require distinguishing chemical signatures such as that provided by lithium. Theoretical models predict that lithium can be synthesised in AGB stars, whereas no net Li production is expected from other candidates.  It has been shown that in order to reproduce the abundance pattern found in M4, Li production must occur within the polluters, favouring the AGB scenario. Here we present Li and Al abundances for a large sample of RGB stars in M12 and M5. These clusters have a very similar metallicity, whilst demonstrating  differences in several cluster properties. 
Our results indicate that the first-generation and second-generation stars share the same Li content in M12; we recover an abundance pattern similar to that observed in M4. In M5 we find a higher degree of complexity and a simple dilution model fails in reproducing the majority of the stellar population. 
In both clusters we require Li production across the different stellar generations, but production seems to have occurred to different extents. We suggest that such a difference might be related to the cluster mass with the Li production being more efficient in less-massive clusters. This is the first time a statistically significant correlation between the Li spread within a GC and its luminosity has been demonstrated. Finally, although Li-producing polluters are required to account for the observed pattern, 
other mechanisms, such as MS depletion, might have played a role in contributing to the Li internal 
variation, though at relatively low level.
\end{abstract}


\section{Introduction} \label{sec:intro}

No longer considered simple stellar populations, globular clusters (GCs) are at the crux of several unresolved problems in stellar astrophysics. 
Accurate photometric surveys have revealed that some clusters display multiple evolutionary sequences, 
from the main-sequence (MS), to the sub-giant 
and red-giant branches (SGB and RGB, see e.g., \citealt{bedin04}; \citealt{anderson09}; \citealt{piotto12}; \citealt{milone12}). 
The simplest interpretation is that there are (at least) two stellar generations present, 
slightly separated in age and characterised by a different chemical composition (see, however, \citealt{bastian13} for a different view).
Although this result is now clear through clever use of multi-filter photometry,   
the seeds for the paradigm shift were planted by high-resolution spectroscopy and detailed abundance determinations across a large sample of GCs
(e.g., \citealt{kraft94}; \citealt{gratton04}; \citealt{carretta09a}, \citeyear{carretta09b}, \citeyear{carretta11}, \citeyear{carretta13}; \citealt{marino08}, \citeyear{marino11}; \citealt{jp10}).

In contrast to field stars and open clusters (e.g., \citealt{gratton00}; \citealt{desilva09}; \citealt{bragaglia12}), 
archetypal GCs exhibit internal variations in elements 
affected by proton-capture processes (hereafter p-capture elements, such as e.g., C, N, O, F, Na, Mg, Al), 
but remain homogeneous in iron-peak, heavy $\alpha$
(Ca, Ti), and neutron-capture elements (e.g., \citealt{james04}; \citealt{smith08}; \citealt{dorazi10a}).\footnote{This picture is further complicated by the presence in the large GC family of some peculiar cases (e.g., $\omega$~Centauri, M 22, NGC 1851, M15, Terzan 5, NGC 2419), where variations
in metallicity and/or heavy-elements are detected (though at a different extent).}
 
The burning patterns give rise to the well known light-element anticorrelations  (C-N, O-Na, Mg-Al), with the Na-O anticorrelation suggested as the GC's defining feature (\citealt{carretta10}). 
The fact that these chemical peculiarities are detected also in un-evolved or scarcely evolved GC stars 
(\citealt{gratton01}; \citealt{rc02}) implies that a fraction of first-generation (FG) stars have 
simultaneously activated CNO, NeNa, and (possibly) MgAl cycles in their interiors in order to deplete C, O and Mg and enhance N, Na and Al, respectively.
From the ejecta of these progenitors, the second-generation (SG) stars were born and currently account for the majority of the stellar population in GCs 
(about 2/3, see \citealt{carretta10}).
Unfortunately, the nature of the progenitors responsible for the internal chemical enrichment remains unclear. 
Intermediate-mass asymptotic giant branch stars (IM-AGB, \citealt{cottrell81}; \citealt{ventura01}) and fast rotating massive stars 
(FRMS, \citealt{decressin07})  remain the prime candidates, although neither offer a satisfactory explanation (we refer to \citealt{gratton12} for an updated and comprehensive review on the multiple population framework).

Besides the imprinted abundance patterns in GCs, lighter nuclei such as \el{7}{Li}, C and N display evidence of \textit{in situ} processing; 
their abundances change as a function of RGB luminosity (beyond the extent of the first-dredge up, FDU). This is not predicted by standard stellar theory and is one example of the need 
for `extra mixing' in numerical models. 

Surveys of C and N in GCs have been used extensively to study the RGB extra-mixing problem 
\citep{2003ApJ...593..509D, 2011ApJ...728...79A,2012ApJ...749..128A} but the recent availability of Li provides a complementary and very powerful diagnostic.  
By virtue of its fragility, Li is a sensitive probe of mixing in stars. It is produced during H burning when \el{7}{Be} 
captures an electron as part of the $pp$~{\sc ii} chain. At these temperatures (T $\gtrsim$ 2 MK), it is also highly favourable for \el{7}{Li} to subsequently capture a proton to produce two \el{4}{He} nuclei. As first pointed out by \citet{1971ApJ...164..111C}, efficient mixing in the burning region can transport the material rich in \el{7}{Be} to cooler temperatures where a further proton capture is avoided. This is the so called {\em ``Cameron-Fowler \el{7}{Be} transport mechanism''}. 
Conversely, \el{7}{Li} is easily destroyed whenever material is transported from the cool surface to the interior of the star. The surface Li abundance serves as a key indicator for either of these internal processes. 

Lithium finds itself at the centre of another long standing discrepancy, that is  
its abundance, as measured in Population~{\sc ii} dwarfs, is factors of 2-3 lower than that predicted by 
Big Bang nucleosynthesis (e.g., \citealt{2008JCAP...11..012C}). 
Currently, the favoured explanation for this inconsistency is that the stars themselves are responsible for the depletion via a different kind of process driven by a different mechanism (such as atomic diffusion, i.e., the transport of chemicals due to temperature, pressure and abundance gradients; see e.g., \citealt{2005ApJ...619..538R}). To date, this cosmological problem has served as the prime motivation for obtaining Li abundances in GC dwarfs (\citealt{pasquini05}; \citealt{bonifacio07}; \citealt{lind09}; \citealt{gonzalez09}; \citealt{mucciarelli11}, \citeyear{mucciarelli12} and references therein). 

\nocite{dm10} D'Orazi \& Marino (2010, hereafter DM10) examined the Li abundances in the mildly metal-poor GC NGC 6121 (M4, [Fe/H]=$-$1.16, \citealt{harris96}, updated in 2010), but focussed on giant stars (104 RGB stars with 32 targets located below the bump luminosity).
The shift in scientific motivation encouraged a shift in target selection: rather than investigate the Spite Plateau (\citealt{spite82}) and hence un-evolved stars, the purpose of that study was twofold: 
\begin{itemize}

\item[*] Constrain the nature of the first (polluting) generation of stars in the cluster. At the high temperatures at which the CNO occurs
(T$\gtrsim$20 MK) it is expected that Li is completely destroyed (T$_{\rm burning}\approx$~2.5 MK). Thus, along with displaying the O-Na, C-N, and possibly Mg-Al  anticorrelations, the multiple population scenario predicts that Li and O/C/Mg should be positively correlated, while Li and Na/N/Al anticorrelated
(if no Li is produced by the polluters). The FG stars should be Li rich, whereas stars formed 
from the ejecta processed at extremely high temperatures should be Li poor.


\item[*] Under the assumption that there is no Li production within the polluters, shed light on the dilution process within this GC and try to determine the amount of pristine (and of polluted) material present in each star. 
\end{itemize}  

For both these goals, any stars not yet experiencing extra mixing are suitable targets. 
Although Li is depleted by a factor of $\approx 20$ during FDU, the abundance change is simply considered a zero-point offset because FG and SG stars are affected in the same way\footnote{Models in an upcoming paper (Angelou et al. 2014) demonstrate that such an assumption is plausible.  Calculations of FG stars (M=0.80 M$_\odot$, Y=0.24) and an extreme population  (M=0.80 M$_\odot$ and Y=0.40) exhibit a  a difference in A(Li)$\sim$ 0.1 dex after FDU, on account of a slightly deeper penetration of the convective envelope.  We expect that if a star is still visible on the giant branch and has such an extreme helium abundance it must necessarily be less massive ( M $\lesssim$ 0.65 M$_\odot$). In such a star the difference in A(Li) is smaller than 0.1 dex after FDU. Even by selecting an unreasonably extreme mass and composition the models predict a difference in Li within the observational uncertainty.}. 
Thus, DM10 could afford to turn their attention to brighter targets, provided the stars were situated below the bump in the luminosity function of the RGB (LF bump). Their main finding was the lack of a Li-Na anticorrelation in M4 with the FG (Na poor, A(Li)=1.34$\pm$0.04) 
and SG (Na rich, A(Li)=1.38$\pm$0.04) sharing the same Li abundance. The implications are that Li has been produced between the different stellar generations and, crucially, this abundance pattern is not an outcome of dilution processes with primordial material. 
This striking result requires the progenitors to produce Li. An outcome which is not currently predicted by massive star evolution \citep{decressin07} nor massive binaries \citep{demink09}.
DM10 therefore provided, for the first time, strong observational evidence that IM-AGB stars seem to play a significant role in the internal enrichment of GCs (at least in this system).

In this study we expand upon previous results and present Li abundances in the GCs NGC 6218 (M12) and NGC 5904 (M5).
Our approach, which focuses on stars brighter than the turn-off, allows us to target more distant systems, whereas previous studies
were forced to analyse the near-by GCs (M4, NGC 6397 and 47 Tuc).
M12 and M5 were specifically chosen to improve our understanding of Li across the GC mass and metallicity distributions.
Both clusters are similar in metallicity (and they are similar to M4) but they differ significantly in mass. 
We note that in NGC~6218, \cite{carretta07} detected that stars brighter than the LF bump possess statistically higher Na than those below. 
This is an expected result from the presence of two populations with distinct He abundances and hence different bump luminosities (\citealt{salaris06}). 

This work is part of a long-term project aimed at determining homogeneous Li abundances at all RGB luminosities (hundreds of stars) in a large number of GCs. Clusters covering a range in mass, metallicity, HB morphology and shape/extent of the Na-O anticorrelation are required to probe the relationship between the Li abundance 
and the GC global parameters. Besides the scientific motivation addressed in this study, such a data set will also provide stringent constraints on stellar evolution and mixing processes in stars.

The paper is organised as follows: in Section~\ref{sec:obs} we provide information on sample, data reduction and abundance analysis; 
our results are presented and discussed in Section~\ref{sec:results}. A summary of our findings concludes the manuscript (Section~\ref{sec:summary}).


\section{Observations, data reduction and analysis}\label{sec:obs}
\begin{figure*} 
\centering
 \includegraphics{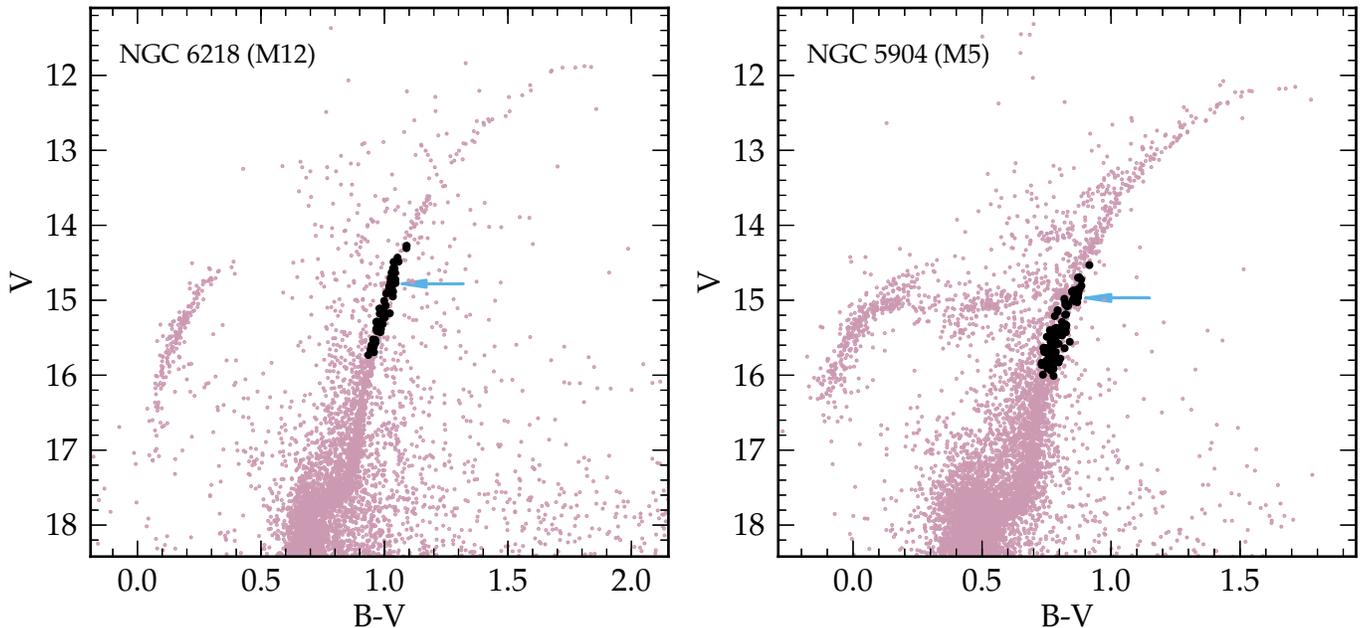}
\caption{The colour-magnitude diagrams for the GCs M12 and M5 (left and right-hand panels, respectively). Stars targeted during this survey are emboldened.
The arrows indicate the location of the LF bump from \cite{nataf13}.}
 \label{fig:cmd}
\end{figure*}

We utilised the multi-object FLAMES@VLT facility  (\citealt{pasquini02}) to collect intermediate-resolution 
spectra of RGB stars, both below and above the bump luminosity, in our target clusters (Program: 087.2-0276(A), PI: VD). 
Employing the HR15N setup (6470$-$6790~\AA), our wavelength coverage included the Li~{\sc i} doublet at 6707.78\AA\ with a nominal resolution of 
R=17,000. We observed a total of 72 stars in NGC 6218 and 113 stars in NGC 5904 using this configuration. 
The sample was selected from the photometric catalogues provided by Momany et al. (private communication). We imposed that targets   
lack companions within 3$^{"}$ or with companions but not closer than 2$^{"}$ and fainter than 2 magnitudes.
We refer to \citeauthor{momany03} (\citeyear{momany03}, \citeyear{momany04}) for details about photometric data reduction and analysis. 
The colour-magnitude diagrams for both GCs are shown in Fig.~\ref{fig:cmd}, with target stars emboldened.
The spectroscopic data reduction was performed by the ESO personnel through the dedicated software that produces 
extracted, bias subtracted, flat-field corrected and wavelength calibrated spectra. 
In addition, continuum normalisation, radial velocity computation, shift to rest-frame and 
combination of multiple exposures were all carried out within IRAF\footnote{IRAF is the Image Reduction and Analysis Facility, a general purpose software system for the reduction and analysis of astronomical data. IRAF is written and supported by National Optical Astronomy
Observatories (NOAO) in Tucson, Arizona.}. The typical S/N ratios (per pixel) of our target stars range from 
60 to 150 at 6700\AA.

From the original sample, we discarded 4 stars in NGC 6218 and 6 stars in NGC 5904 as their radial velocities were 
more than 3$\sigma$ away from the respective cluster mean. We derived heliocentric radial velocities of v$_{rad}= -43.31 \pm 0.41$ km~s$^{-1}$ (r.m.s $=3.35$ km s$^{-1}$, 68 stars for NGC 6218) and 
v$_{rad} =53.05 \pm 0.51$ km s$^{-1}$ (r.m.s $=5.22$ km~s$^{-1}$, 107 stars NGC 5904), 
which are in reasonable agreement, in view of the zero point uncertainties, with values published by \cite
{harris96}, that is v$_{rad}=- 41.4 \pm 0.2 $ km~s$^{-1}$ and v$_{rad} =53.2 \pm 0.4$ km~s$^{-1}$.

The stellar atmospheric parameters for our sample stars were derived in the following way. 
We first calculated initial \Teff\ values from ($V-K$) colours (with $V$ from our photometry and 2MASS $K$ magnitudes, \citealt{skru06}) 
and the calibration by \cite{alonso99}; metallicity and reddening were retrieved from the Harris' catalogue, that is [Fe/H] $=-1.33$ and 
E($B-V$) $=0.19$ for NGC 6218, and [Fe/H] $=-1.29$ and E($B-V$) $=0.03$ for NGC 5904.
The reddening values were converted to E($V-K$) via the relationship by 
\cite{cardelli89}, i.e., E($V-K$) $=2.75 \ \times$ E($B-V$).
Our final adopted temperatures come from a relationship between those \Teff\ values and $V$ magnitudes, following the same approach as in 
\citeauthor{carretta06}, (\citeyear{carretta06}; \citeyear{carretta09a}).
To derive this relation, we used only ``well-behaved''
stars in each cluster, that is stars with magnitudes in both visual
and infrared filters (keeping only high-quality flagged 2MASS photometry), and lying on the RGB. 
Comparing the stars in common between the present study and \citeauthor{carretta07} (\citeyear{carretta07}, \citeyear{carretta09a})
we obtained a difference (in the sense ours $minus$ Carretta) of $\Delta$\Teff $=-52 \pm 8$ K and $\Delta$\Teff $=5 \pm 2$ for NGC 6218 and NGC 5904, respectively.
The larger difference for NGC 6218 is likely related to the different photometric catalogue adopted by \cite{carretta07}. 
Surface gravities ($\log$$g$) were then derived by assuming our final \Teff~values, distance moduli of ($m-M$)$_V=14.04$ and 
($m-M$)$_V=14.46$ (\citealt{harris96}), 
a bolometric solar magnitude of M$_{bol, \odot} = 4.75$, and masses of 0.85 M$_{\odot}$, with the standard formula:
$$\log\frac{g}{g_\odot}=\log\frac{M}{M_\odot}-\log\frac{L}{L_\odot}+4\log\frac{T_{\rm eff}}{T_{\rm eff,\odot}}$$
Finally, microturbulence velocities ($\xi$) were computed from the relationship by \cite{gratton96}: $\xi =2.22 - 0.322 \times \log \ g$, while the input metallicity was taken from \cite{harris96}. 

The Li abundances were inferred via spectral synthesis with the driver $synth$ in {\sc MOOG} (\citealt{sneden73}, 2013 version) 
and stellar atmospheres by \cite{ck04}, with $\alpha$-enhancement (+0.4 dex) and no overshooting\footnote{Available at ~\url{http://kurucz.harvard.edu/grids.html}}.
We investigated the impact of this choice by deriving Li abundances for all our sample stars using three different sets of 
atmospheric models, namely the Kurucz (\citeyear{kur93}) grid with and without overshooting and the Castelli \& Kurucz (\citeyear{ck04}) with solar-scaled composition (i.e., no $\alpha$-enhancement). The differences in the derived Li abundances are always smaller than 0.05 dex and can be safely considered as negligible. 
Adopting the same line lists as in DM10 (see \citealt{dorazi10b} and DM10 for details on atomic parameters), we computed a grid of synthetic spectra for each star by varying the Li abundances until the best match between observed and synthetic profiles was attained.   
The synthetic spectra were calculated covering a wavelength range from 6695 \AA~to 6722 \AA, exploiting the strong Ca~{\sc i} line at 6717.69\AA~to evaluate the spectral broadening (which was assumed to be Gaussian).
We were able to determine the Li abundances for 63 stars for NGC 6218 and 99 stars for NGC 5904, 
because of the occurrence of cosmic rays and/or due to lower S/N ratios that hampered measurements in 5 and 8 stars, respectively for each GC.

\begin{figure*} 
\centering
\includegraphics[width=18cm]{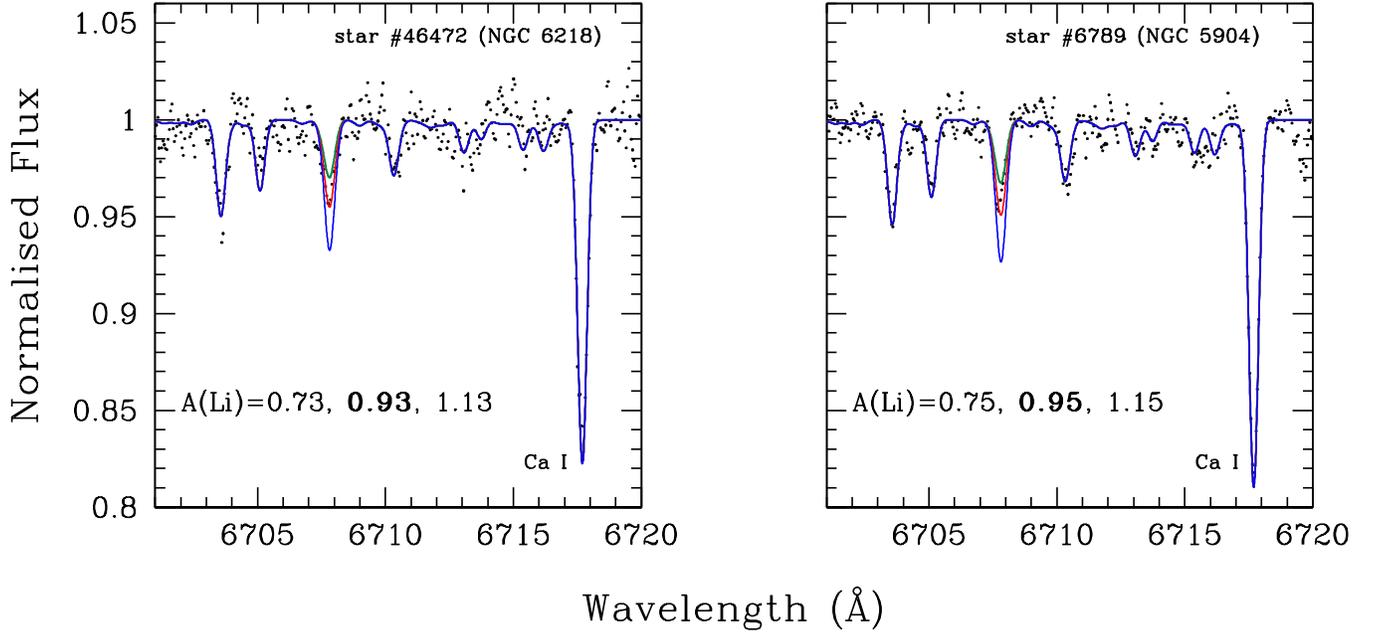}
\caption{Example of the spectral synthesis for the Li~{\sc i} line for stars \#46472 (NGC 6218) and \#6789 (NGC 5904).}
\label{fig:synli}
\end{figure*}

\begin{figure*} 
\centering
\includegraphics[width=18cm]{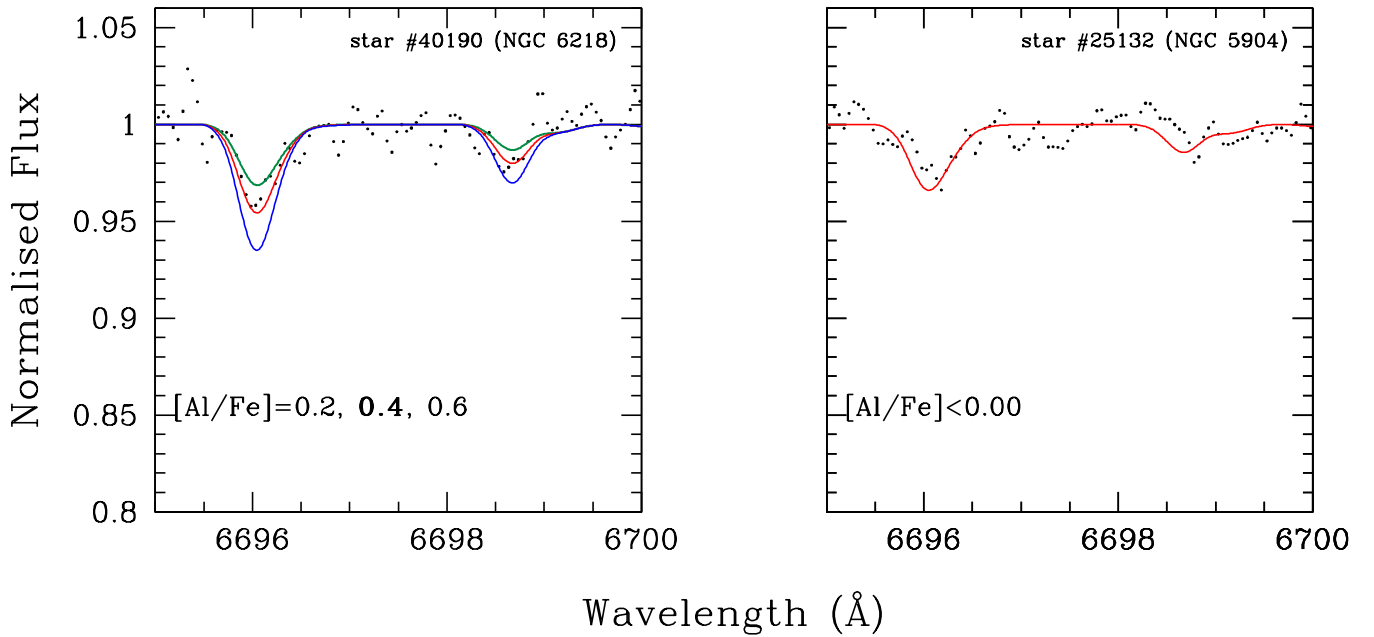}
\caption{Example of the spectral synthesis for the Al~{\sc i} lines for stars \#40190 (NGC 6218) and \#25132 (NGC 5904; here only an upper limit is given).}
\label{fig:synal}
\end{figure*}

The spectral coverage guaranteed by the HR15N grating, although not capturing any suitable features for Na and O abundance determinations, 
allowed us to investigate the Al content by synthesising the Al~{\sc i} 
doublet at 6696\AA\ and 6698\AA. The adopted atomic parameters for those lines were: log $gf = - 1.35$ and log $gf =-1.65$, for the 6696\AA\ and 6698\AA\ features, respectively. In NGC 6218 we could obtain Al abundances for a sample of 54 out of 63 stars, 
whereas in NGC 5904 we analysed 93 stars (out of 99), with 61 detections and 32 upper limits.
Examples of spectral synthesis for Li and Al are given in Figures~\ref{fig:synli} and \ref{fig:synal} for stars in both clusters.


\subsection{Error budget}
Two kind of uncertainties affect our derived abundances, that is internal (star-to-star) and systematic (cluster) errors.
The main aim of our paper is to search for (possible) evidence of spreads in Li and Al abundances, thus we focus 
on the first source of errors.
\begin{table}
\centering
\caption{Sensitivities of abundances to atmospheric parameters for star \#40129.}\label{t:sensi}
\begin{tabular}{lccr} 
 \hline \hline
Species        & \Teff+100 & log$g$+0.2 & $\xi$+0.1 \\
\hline
$\Delta$A(Li)    &  0.10     &   0.02     &    0.00   \\
$\Delta[$Al/Fe$]$&  0.07     &   0.06     &    0.02   \\
\hline\hline
\end{tabular}
\end{table}
The internal errors are mainly related to $(i)$ the best-fit determination of synthetic spectra with observed spectra (which is in the range 0.07 - 0.10 for our target stars and reflect
uncertainties in the continuum placement and S/N ratios of the spectra) and $(ii)$ to the atmospheric parameters, i.e., \Teff\, $\log \ g$, and microturbulence velocities $\xi$ (the adopted metallicity [A/H] in the model atmosphere has a negligible impact).
In order to assess the contribution related to the stellar parameters we first need to ascertain the sensitivities 
of our species to changes in 
atmospheric quantities (the partial derivatives in Equation~\ref{eq:sensi}). 
To do this we proceeded in the standard way, that is by varying one parameter at the time and 
inspecting the corresponding change in the resulting abundance 
(see Table~\ref{t:sensi} where sensitivities are reported for one sample star with median \Teff). 
The following step is to evaluate the actual error in atmospheric parameters (i.e., $\sigma_{T_{\rm eff}}$,  $\sigma_{logg}$, $\sigma_\xi$).
The $\sigma_{T_{\rm eff}}$ can be estimated from the error on the slope of the relation between initial \Teff\ values (from ($V-K$) colours) and 
$V$ magnitudes, which result in 18K for both GCs. 
Errors in $\xi$ instead come from the scatter around the relationship of $\xi$ $vs$ log $g$ 
by \citeauthor{gratton96} (\citeyear{gratton96}, that is 0.2 km s$^{-1}$). 
Finally, the $\sigma_{log g}$ contains different terms due to the uncertainties in stellar masses (which is, however, less than 
$\approx$10\% of the mass), errors due to luminosity (in turn related to magnitudes, distance moduli and bolometric corrections) 
and those in temperatures. All these contributions are anyway significantly small and results in internal errors 
in log$g$ values less than 0.05 dex.

The total internal error on a given species is then calculated by summing in quadrature all the different contributions, i.e.: 
\begin{footnotesize}
\begin{equation}\label{eq:sensi}
\sigma= \sqrt{\sigma^2_{best} + 
\left (\frac{\partial \log(\epsilon)}{\partial T_{\rm eff}}\right ) ^2  \sigma_{T_{\rm eff}}^2 +
\left (\frac{\partial \log(\epsilon)}{\partial logg}\right ) ^2  \sigma_{logg}^2 +
\left (\frac{\partial \log(\epsilon)}{\partial \xi}\right ) ^2  \sigma_{\xi}^2}
\end{equation}
\end{footnotesize}

Given the small uncertainties in stellar parameters, the total errors on Li and Al abundances are almost entirely related to the best-fit determination;
typical values range from 0.10$-$0.13 for Li and 0.14$-$0.16 for [Al/Fe].


\section{Results and Discussion}\label{sec:results}
Elemental abundances (Li and [Al/Fe]) are displayed in Tables~\ref{t:m12} and~\ref{t:m5}, where we list the identification number 
for each star and $V$ magnitudes (from the Momany et al. photometry), the S/N ratios at 6700\AA\ along with stellar atmospheric parameters (the complete Tables are made available online only). 
Our results are presented separately for each target cluster in Sections \ref{sec:m12} and \ref{sec:m5} 
for NGC 6218 and NGC 5904, respectively. 
We then provide a general discussion on the Li abundance pattern observed in GCs by summarising 
and discussing findings from this study along with previous investigations (Section~\ref{sec:spreads}).    
\begin{table}
\centering
{\renewcommand{\arraystretch}{1.5}
\renewcommand{\tabcolsep}{0.1cm}
\caption{Stellar parameters, Li and Al abundances for targets in NGC 6218 (M12). This table is available in its entirety in a machine-readable form in the online journal. A portion is shown here for guidance regarding its form and content.}\label{t:m12}
\begin{tabular}{lccccccc} 
  \hline \hline
Star ID & $V$  & S/N & \Teff & $\log \ g$ & $\xi$ & A(Li) & [Al/Fe]  \\
        & (mag) &    & (K)  & (cm s$^{-2}$) & (km s$^{-1}$) & dex & dex \\
  \hline
10219 & 15.28 & 78 & 4946 & 2.59 & 1.39 & 1.08  & ~---  \\
26629 & 15.60 & 60 & 5016 & 2.75 & 1.34 & 1.06  & 0.60 \\
26778 & 14.57 & 115 & 4787 & 2.23 & 1.50 & 0.53  & 0.00  \\
31393 & 15.17 & 100 & 4921 & 2.53 & 1.41 & 0.93  & ~---    \\
31600 & 14.95 & 90 & 4871 & 2.42 & 1.44 & 0.95  & 0.00 \\

\hline
\end{tabular}}
\label{tab:m12results}
\end{table}
\begin{table}
\centering
{\renewcommand{\arraystretch}{1.5}
\renewcommand{\tabcolsep}{0.1cm}
\caption{Stellar parameters, Li and Al abundances for targets in NGC 5904 (M5). Entries with asterisks indicate upper limits.  This table is available in its entirety in a machine-readable form in the online journal. A portion is shown here for guidance regarding its form and content. 
}\label{t:m5}
\begin{tabular}{lccccccc} 

  \hline \hline
Star ID & $V$ & S/N & \Teff & $\log \ g$ & $\xi$ & A(Li)  & [Al/Fe]  \\
        & (mag)&    & (K)  & (cm s$^{-2}$) & (km s$^{-1}$) & dex & dex \\
  \hline
229  & 15.63 & 100   & 5000 & 2.57 & 1.39	& 1.02        &  ~0.00$^*$  \\
394  & 15.13 & 130   & 4888 & 2.32 & 1.47	& ~0.65$^*$    &  0.35  \\
1069 & 15.91 &  95   & 5063 & 2.71 & 1.35	& 1.05        &  ~0.10$^*$  \\
1476 & 15.03 & 140   & 4865 & 2.27 & 1.49	& 0.97        &  ~0.00$^*$     \\
1778 & 15.56 & 115   & 4983 & 2.53 & 1.40	& 1.00        &  0.10   \\

\hline
\end{tabular}}
\label{tab:m5results}
\end{table}

\subsection{NGC 6218 (M12)}\label{sec:m12}

\begin{figure} 
 \includegraphics[width=8cm]{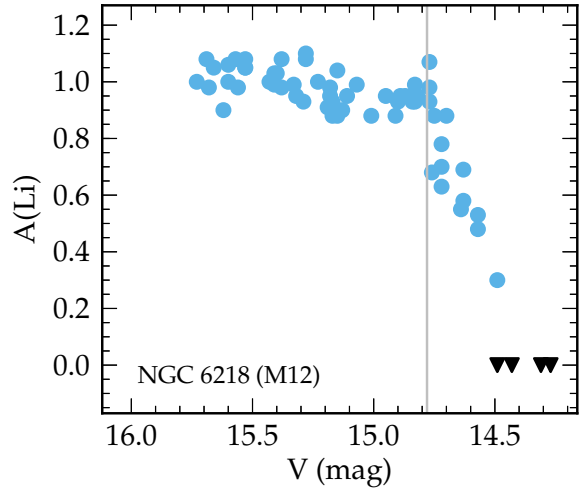}
\caption{A(Li) as a function of visual magnitude in the globular cluster NGC 6218 (M12). Blue circles denote lithium detections whereas black triangles represent upper limits. The solid grey line indicates the magnitude of the LF bump in this cluster, V$=14.78$, as given by \cite{nataf13}.}
 \label{fig:M12LiVm}
\end{figure}

In Figure \ref{fig:M12LiVm}, A(Li) is plotted as a function of $V$ magnitude for all our sample stars in NGC 6218. Here blue circles denote detections, black triangles 
upper limits and the solid grey line indicates the magnitude of the LF bump, located at $V$=14.78  according to \citeauthor{nataf13}~(\citeyear{nataf13}). 
The magnitude of the LF bump is generally believed to denote the beginning of extra mixing in these low-mass stars: there is excellent agreement between photometry and spectroscopy
regarding the onset of this event, with stars exhibiting a declining trend of Li abundances as a function of luminosity once the bump is reached. 
The brightest stars in our sample have had their lithium content significantly depleted and only upper limits can be provided. 
Moreover, a slightly decreasing trend of Li with magnitudes is present for our targets with V ranging between V=15.75 and the bump: 
the correlation coefficient is found to be $r$=0.55, which is significant at more than 99.9 \% level (the probability that this happened by chance is less than 0.1 \%). The slope of the correlation is 0.13 +/- 0.03, which is very close to the accuracy imposed by
our observational uncertainties ($\sim$ 0.10-0.13 dex). Given that our errors are likely overestimated, we are tempted to conclude that this Li pattern is real. The trend is clearly not related to the multiple population scenario, because it is present in both FG and SG stars (divided according to the Na and O abundances, as in \citealt{carretta09a}). As pointed out by the referee, there could be two possible explanations for this trend: ($i$) in situ depletion as stars evolve along the sub giant branch (contrary to standard theory); or ($ii$) increased previous Li depletion as a function of mass along the sub giant branch, possibly showing signs of a Li dip in metal-poor stars analogous to the Population~{\sc i} F dwarf dip.

\begin{figure*} 
\centering
 \includegraphics[width=12cm]{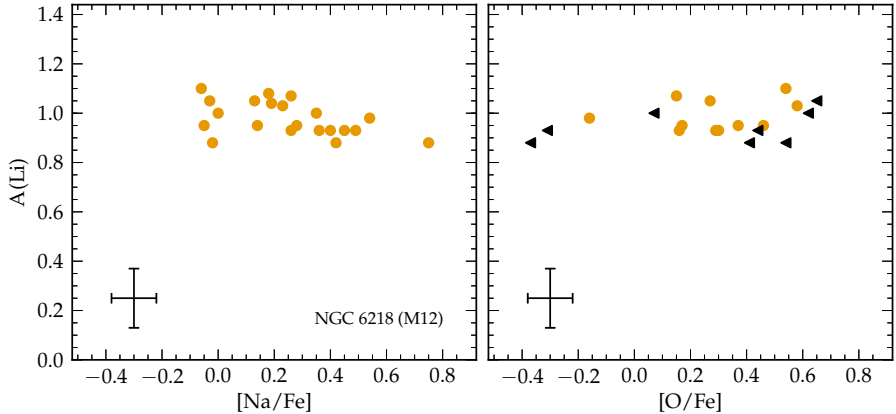}
\caption{Left Panel: A(Li) for stars below the RGB bump as a function of [Na/Fe] and [O/Fe] as determined by \cite{carretta07}. 
Orange circles denote detections and triangles upper limits in O abundances. Error bars indicate the typical internal error.}
 \label{fig:LINaOM12}
\end{figure*}

\begin{figure} 
 \includegraphics[width=8cm]{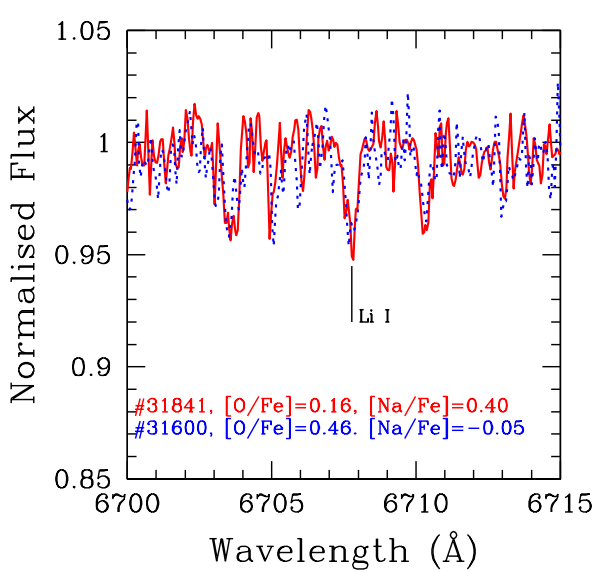}
\caption{Comparison of the spectra for two stars in M12 with very similar stellar parameters and Li abundances, but very different Na and O content (see text for discussion).}
 \label{fig:spectra_m12}
\end{figure}

By considering only giants fainter than the RGB bump, we find a mean Li abundance of A(Li) $=0.98 \pm 0.01$ (r.m.s. 0.06, 44 stars). 
{\it This is consistent with no Li variation in this cluster} (the fact that the standard deviation is formally lower than observational 
uncertainties -0.10/0.13 dex- indicates that the measurement errors are probably over-estimated).
It is implicit that any spread we detect is not the true internal (intrinsic) Li dispersion. There is a component that is due to observational errors. Similarly, in clusters where we state that the dispersion is consistent with no lithium variation, i.e., M4 and M12, there may in fact be a small spread below the measurement uncertainties. We cannot prove zero dispersion, but only provide statistical limits to its size. 
Given our observational errors, we obtain that 
the internal dispersion in Li, estimated using the standard deviation from the mean, cannot be larger than about $\sim$ 0.1 dex.
The constancy of Li abundances is a noteworthy result because NGC 6218 is known to display, along with the large majority of Galactic GCs, 
large variations in p-capture elements. 
As previously mentioned in Section~\ref{sec:obs}, no information on Na and O abundances can be gathered from our spectra. However, we have
stars in common with the survey by \cite{carretta07}: out of 44 stars (not yet experiencing $in~situ$ extra mixing), we have Na and O abundances for 21 and 18 stars, respectively. 
In Figure \ref{fig:LINaOM12} we show our A(Li) against their [Na/Fe] and [O/Fe] ratios for those stars in common: variations of almost 1 dex in Na and more than 1.2 dex in O do not coincide with changes in the Li abundance. 
There seems to be the hint of a weak anticorrelation between Li and Na abundances, however the Pearson's correlation coefficient results in $r$=$-$0.38
(16 degrees of freedom) and is not statistically meaningful: there is a probability larger than 10\% that this correlation could happen by chance.
Following the definition introduced by \cite{carretta09a}, we can group our stars into their respective generation based on their Na content, with 
FG stars defined as having Na abundance of [Na/Fe] $\le$ [Na/Fe]$_{\rm{min}}+0.3$ dex.   
We find that FG stars have an average Li abundance of A(Li) $=1.00 \pm 0.04$ (r.m.s $=0.09$), whilst in SG stars A(Li) $=0.98 \pm 0.02$ (r.m.s $=0.06$). 
Thus, the different stellar populations identified according to their Na abundances are indistinguishable in terms of their Li content. 
In Figure~\ref{fig:spectra_m12} we compare the spectra of two stars with very similar parameters ($\Delta$\Teff $=37K$), but differences in Na and O abundances of more than a factor of 2. It is evident that there is no remarkable difference in the Li~{\sc i} line strength, entailing that those stars have to share a very similar Li content (virtually the same within the observational uncertainties).

Our spectral coverage also permitted us to obtain Al abundances for a sub-sample of 54 stars. 
We derived a mean Al abundance of [Al/Fe] $=+0.21 \pm 0.03$ (r.m.s $=0.19$), which is in excellent agreement with values published 
by \cite{carretta09b} ([Al/Fe] $=0.20 \pm 0.05$, r.m.s $=0.18$) based on a sample of 11 bright giants observed with the high-resolution UVES spectrograph. 
\cite{johnson06} analysed intermediate-resolution spectra (R$\approx$15000) for 21 RGB stars in this GC, deriving stellar parameters, 
metallicity, p-capture and n-capture element abundances. They obtained an average [Al/H] $=-1.00 \pm 0.03$ compared to our value of [Al/H] $=-1.16 \pm 0.03$ (we directly compare [Al/H] because there is an offset in metallicity of about 0.2 dex between the two studies). Taking into account the measurement uncertainties, and a difference in the log $gf$ for the Al doublet at 6696$-$6698 \AA~of 0.20 and 0.24 (in the sense ours $minus$ theirs), the two mean values agree very well.

\begin{figure}
\centering 
 \includegraphics[width=8cm]{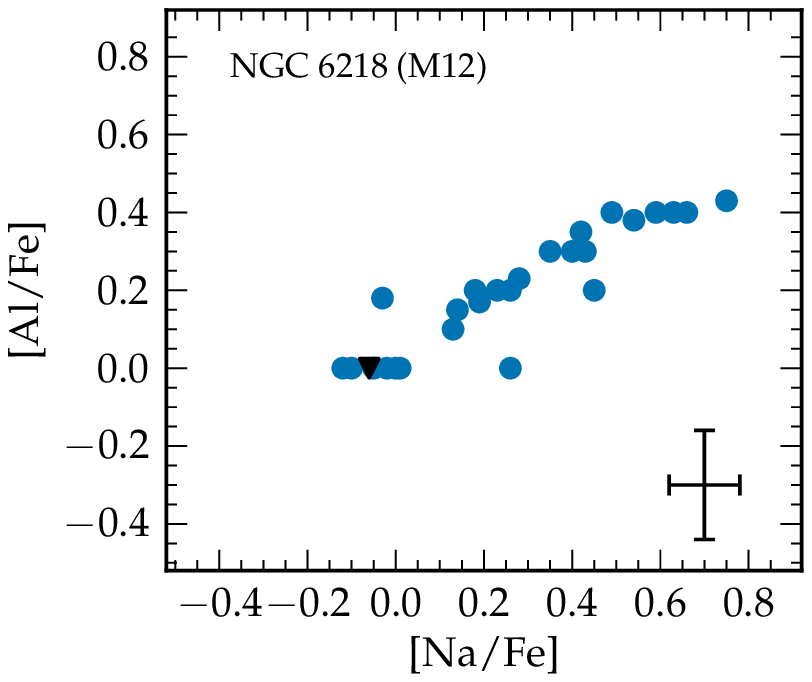}
\caption{[Al/Fe] from the present study as a function of the [Na/Fe] derived by \cite{carretta07}}\label{fig:naal_m12}
\end{figure}

Our results confirm that the Al content does vary in this GC and that Al and Na abundances are positively correlated, as expected from the activation of NeNa and MgAl cycles. This is demonstrated in Figure~\ref{fig:naal_m12}, where our [Al/Fe] ratios are plotted as a function of [Na/Fe] from 
\citeauthor{carretta07} for stars in common.
Our sample stars span a range of $\approx$ 0.8 dex in [Al/Fe] (peak-to-peak variation), 
very similar to the value found by \cite{carretta09b} from UVES spectra (i.e., $\Delta$[Al/Fe]~$\approx$ 0.7 dex), 
whereas this is larger than what has been found by \cite{johnson06}, $\Delta$[Al/Fe]$\approx$ 0.4 dex. 
It is not straightforward to determine the cause of such discrepancy, and statistics may certainly play a role (our sample is a factor of 2 larger). 
However, we note that according to \cite{johnson06} there are no Al-poor stars (FG) in their sample. The minimum value in this population  
being [Al/Fe] $=0.35$ dex
(while the maximum Al abundance is in good agreement with our value as well as with those by \citealt{carretta09b}, i.e., roughly at $\approx$ 0.7 dex level).

\begin{figure} 
\centering
 \includegraphics[width=8cm]{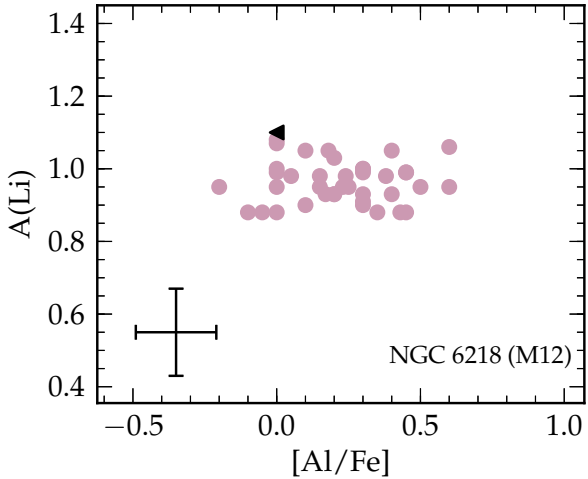}
\caption{A(Li) for stars below the RGB bump as a function of [Al/Fe] abundance in the globular cluster NGC 6218 (M12). Lavender circles denote lithium detections whereas the black triangle represents an upper limit to the [Al/Fe] abundance. The typical internal error for the abundances are indicated by the error bar.}
 \label{fig:M12LiAl}
\end{figure}

In Figure \ref{fig:M12LiAl} we plot A(Li) as a function of [Al/Fe] abundance, where lavender circles denote detections and the black triangle indicates an upper limit to the [Al/Fe] ratio. If we compare the variation in Al from our entire sample ($\approx$0.8 dex) to that we determined from common stars with  \citeauthor{carretta09b} (\citeyear{carretta09b}, see Figure~\ref{fig:naal_m12}, $\approx$0.4 dex) there is almost a factor of two difference.  This might be a mere statistical effect. However, it could be that, since stars in common between the two works are brighter, the scatter increases as luminosity decreases (because the spectral lines get weaker at higher temperatures). We checked the presence of possible trends between our [Al/Fe] ratios and the $V$ magitudes and we concluded that there is a slight increase in the Al dispersion at lower luminosity but the trend is weak (the effect is anyway well within the observational uncertainties).

Analogously to what is revealed in the Li-O and Li-Na planes, 
while Al displays a large variation among our sample stars, the Li abundance remains constant.   
{\it The large variations in all the
p-capture elements under scrutiny here are not accompanied by analogous changes in Li.} 
The implication is that Li production must have occurred across the different stellar generations, ruling out a major contribution by 
FRMS to the GC internal enrichment, and favouring the IM-AGB candidate. 
In this regard, M12 is ``{\it M4-like}'' in its behaviour; we recall that both clusters are of similar (current) mass and metallicity
(see Section~\ref{sec:spreads} for further discussion).  
%
%
%
\subsection{NGC 5904 (M5)}\label{sec:m5}

\begin{figure} 
 \includegraphics[width=8cm]{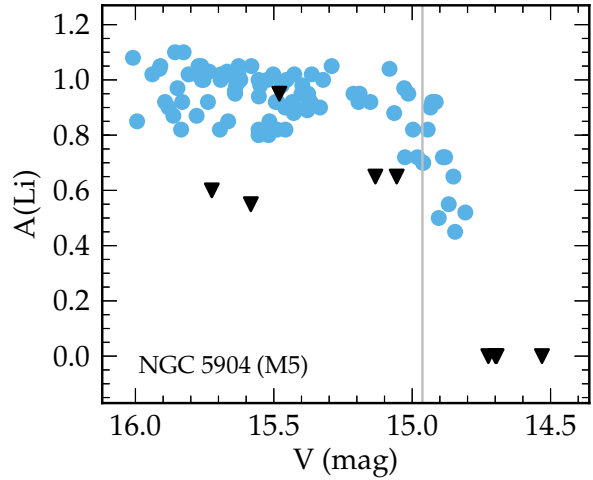}
\caption{A(Li) as a function of visual magnitude in the globular cluster NGC 5904 (M5). Symbols are as for Figure~\ref{fig:M12LiVm}. The solid grey line indicates the magnitude of the LF bump at V$=14.96$ as given by \cite{nataf13}.}
 \label{fig:M5Liv}
\end{figure}

We obtained Li abundances for 99 stars in the massive GC NGC~5904 (M5, [Fe/H] $=-1.29$ dex); 
once again the magnitude range of our sample included giants beyond the LF bump. 
Our derived Li abundances are shown as a function of the $V$ magnitudes in Figure \ref{fig:M5Liv}, with symbols retaining their meaning from Figure \ref{fig:M12LiVm} (i.e., blue circles denote detections, black triangles denote upper limits and the solid grey line indicates the magnitude of the LF bump). This GC possesses two features worthy of mention. 
First, spectroscopically it is ambiguous as to what magnitude extra mixing begins in this cluster. The photometrically derived LF bump, it could be argued, is located at a $V$ magnitude beyond which stars have already begun to deplete their lithium: compared to the trend observed in Figure~\ref{fig:M12LiVm}, the onset of extra mixing is not as clear as in M12. 
Despite the fact that our photometric system is different from that of  \citet{nataf13}, we determined very similar values for the location of the LF bump.
\cite{nataf13} derived $V_{bump} = 14.96 \pm 0.01$  compared to  $V_{bump} =14.97 \pm 0.04$ from the present work. 
Thus, we can state that no major systematic offsets are present between the two catalogues. On the other hand, 
a small shift to fainter magnitudes of 
$\approx$ 0.1 mag for the bump luminosity would result in an agreement between the photometric and spectroscopic
determined location at which extra mixing begins. Such a small difference in the required magnitude \textit{could} be attributed to observational errors. 
Nevertheless, it will be a point of caution in discussions hereinafter.  
Secondly and most interestingly, two stars in particular (the two upper limits with V magnitude $ > 15.5$ in Figure \ref{fig:M5Liv}) were found to be severely lithium deficient for their evolutionary phase. Both their radial velocity measurements and metallicity suggest that they are indeed members of the cluster but their Li abundances are inconsistent with the post-FDU composition of the other stars. Whether this translates to deeper FDU, some sort of extra mixing, a rare  evolutionary event (although it had to happen at least twice) or it is related to variations in p-capture elements remains unclear (see the following discussion).   
   
\begin{figure*} 
\centering
 \includegraphics[width=12cm]{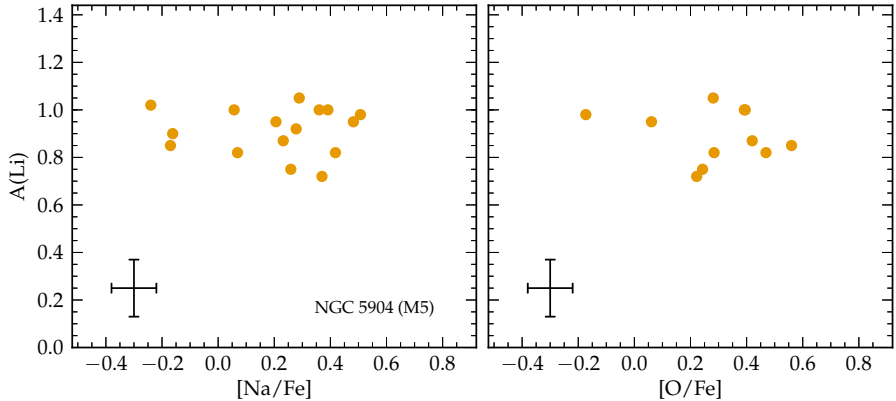}
\caption{Left Panel: A(Li) for stars below the RGB bump as a function of [Na/Fe] as determined by \citet{carretta09a}. Right Panel: A(Li) from this study as a function of [O/Fe] as determined by \cite{carretta09a}. Orange  circles denote stars for which the abundances of both species of interest have been measured and error bars indicate the typical internal error in each ratio.}
 \label{fig:LiNaOM5}
\end{figure*}

\begin{figure}
\centering 
 \includegraphics[width=8cm]{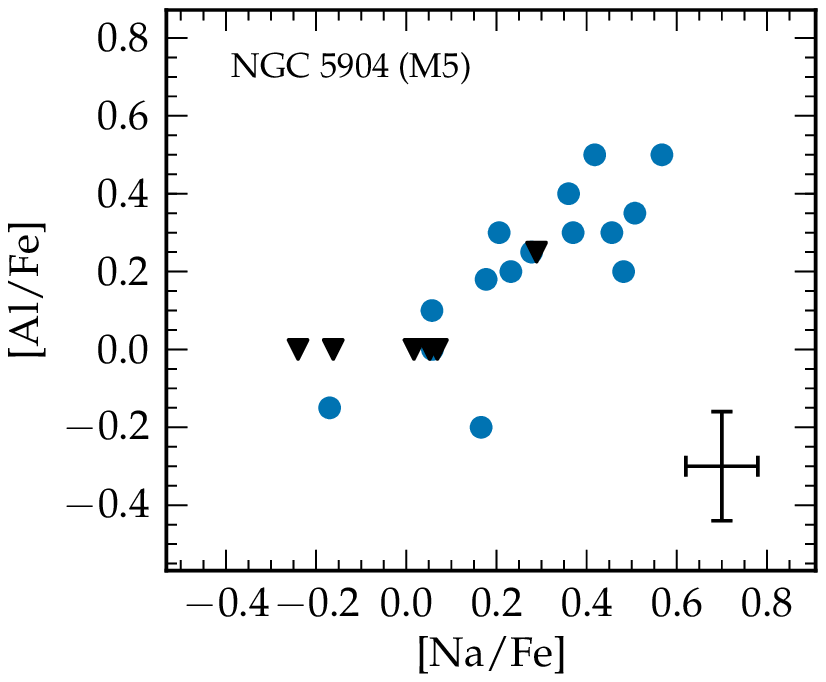}
\caption{[Al/Fe] from the present study as a function of the [Na/Fe] derived by \cite{carretta09a}}
 \label{fig:NaAlm5}
\end{figure}

When considering the 82 stars that are below the magnitude of the LF bump, we find a mean lithium abundance of A(Li) $=0.93 \pm 0.01$ (r.m.s 0.11). 
The standard deviation is roughly of the same order of magnitude of the observational uncertainties, however we need to bear in mind that in the computation of the average, the upper limits in A(Li) are treated as detections. This indicates that the r.m.s is certainly a lower limit to the actual internal 
dispersion in Li abundances. 
Furthermore, as already stated in  Section~\ref{sec:m12}, the measurement errors are quite likely overestimated. 
Unfortunately, we have only 16 and 11 stars for which \cite{carretta09a} have gathered Na and O abundances; 
in Figure~\ref{fig:LiNaOM5} we show the run of Li with [Na/Fe] and [O/Fe] ratios for stars in common 
between the two spectroscopic investigations. There is no evidence for a Li-O positive correlation nor for a Li-Na anticorrelation: 
Na and O extend for $\approx$ 0.7 dex, whilst the Li remain almost constant. 
The only previous determination of Li abundance in this GC is that by \cite{lai11} who derived Li abundances for 
three RGB stars below the RGB bump and found an average of A(Li) $=0.81 \pm 0.06$ (r.m.s $=0.11$). The authors concluded that, given the small size of their sample, they can not comment on the relationship between Li and p-capture elements (e.g., C, Na, O).   
The exiguous number of common stars between this work and \cite{carretta09a} may, however, have prevented us from unveiling the presence of Li variations in conjunction with the other species 
involved in the {\it hot} H-burning, such as Na and O. 
Thus, to get deeper insights into the possible relationship between Li and p-capture elements, we derived the Al abundances for a total of 93 stars.
From our sample we detected a peak-to-peak variation in the [Al/Fe] ratio of $\approx 0.7$ dex, which is the same value found by 
\citet{2001AJ....122.1438I} and is consistent with \citet[][$\approx 0.6$] {shetrone96} and \citet[][$\approx 0.8$ dex]{carretta09b}.   
Considering the sub-group of stars in common with \citeauthor{carretta09a} (\citeyear{carretta09a}, 21 stars), we report our derived Al abundances as a function of the Na determined by  the Carretta et al study in Figure~\ref{fig:NaAlm5}. We obtained a very clear Na-Al correlation, with a Pearson's 
correlation coefficient $r=0.81$, which is significant at more than 99.99\%. In this respect, we confirm results from previous studies 
(see \citealt{carretta09b}, \citealt{2001AJ....122.1438I}, \citealt{shetrone96}).  

\begin{figure}
\centering 
 \includegraphics[width=8cm]{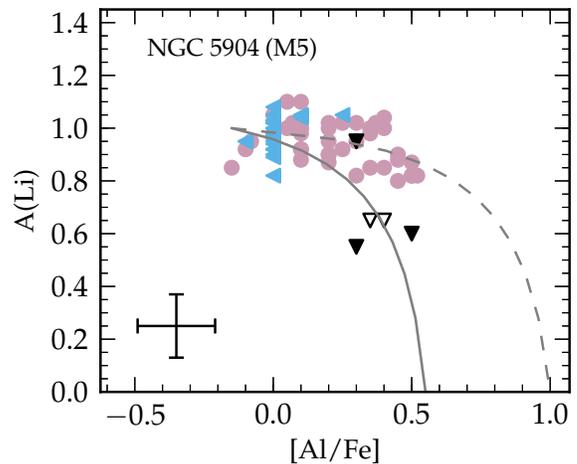}
\caption{A(Li) for stars below the RGB bump as a function of [Al/Fe] abundance in the globular cluster NGC 5904 (M5). Lavender circles denote measurements for both species, black triangles denote upper limits to the lithium abundance and blue triangles denote upper limits to the [Al/Fe] abundance. The two ``peculiar'' stars close to the bump are marked as empty (black) triangles.
The typical internal errors are indicated. The curves represent different dilution models (see text for discussion).}
 \label{fig:M5LiAl}
\end{figure}
\begin{figure} 
 \includegraphics[width=8cm]{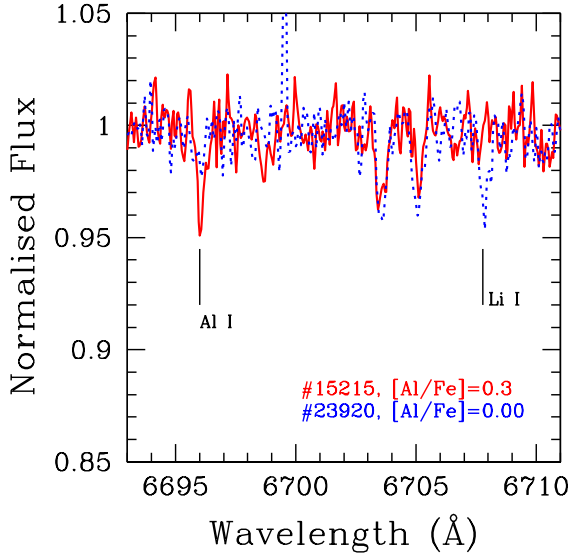}
\caption{Comparison of the spectra for two stars in M5 with different Li and Al abundances (see text for discussion).}\label{fig:spectra_m5}
\end{figure}
As was previously done for NGC 6218, we include a discussion on the [Al/Fe] and Li abundances for those stars in the cluster yet to begin $in~situ$ extra mixing.
By adopting a formal $V_{bump}$ of 14.96, as given from photometry (but keeping in mind possible mismatches between photometry and spectroscopy, as mentioned at the beginning of this Section), 
we plot the derived A(Li) abundances as a function of [Al/Fe] in Figure \ref{fig:M5LiAl}. 
Here lavender circles denote stars where both Li and Al have been measured, black triangles represent stars for which an upper limit to the lithium abundance has been determined and blue triangles represent  upper limits in the derived [Al/Fe] abundance.
There is evidence for a Li-Al anticorrelation, with a Pearson's coefficient r $=-0.44$ (75 degrees of freedom) which is significant at more than 99.9\% 
(the probability that this event can happen by chance is lower than 0.1\%).
If we discard those two stars for which the extra mixing might already have begun (i.e., stars very close to the bump and labelled as 
empty triangles here to emphasise their different behaviour), the anti-correlation is still present (r $=-0.40$). 
Note that even had we discarded the other two upper limits ($V>$15.5), there is still evidence for anticorrelation between Li and Al 
abundances ($r$=$-$0.33, significance level at 99.9 \%) 
However, these other two stars that exhibit Li depletion have magnitudes $V>$15.5, so they are much fainter than the bump. They demonstrate 
the expected abundance pattern if we assume that the processed material that forms the SG stars is Li-poor and Al-rich.
To convince the reader of this possibility, we show in Figure~\ref{fig:spectra_m5} the spectral comparison for one of these stars, namely star \#15215, with another GC member with identical atmospheric parameters (star \#23920, $\Delta$\Teff $=9$K). As is clear from the Figure, \#15215 is Li-poor and (relatively) Al-rich, whereas \#23920 is Li-rich and Al-poor. 
Thus, we might be tempted to conclude that the two (and perhaps the four?) stars that exhibit significant Li depletion (and corresponding Al enhancement) constitute the extreme (E) SG stars in M5. However,  having detected so few of these stars would entail that the fraction of E stars we obtain is about the 3$\pm$1\% of the cluster population, which is lower than the value of 7$\pm$2\% found by \cite{carretta09a} according to their Na and O abundances (see that paper for details on the definition of the PIE groups).

In order to gain a better understanding of the chemical abundance pattern emerging from this study, we determined a dilution model for this GC, as per \citet{2006A&A...458..135P}. 
In this model $[X]$, the logarithmic abundance of species X, is a mixture (given by a dilution factor, \textit{d}) of the original abundance, $[X_o]$, and processed material, $[X_p]$. $[X]$ is determined such that:

\begin{equation}
[X]=\rm{Log}\left[(1-d)10^{[X_o]}+ d  \times 10^{[X_p]}\right].
\end{equation}

In Figure \ref{fig:M5LiAl} we plot as solid line the dilution model with initial abundances of Li=1.00 and [Al/Fe]=$-$0.15 
and processed material having Li=0.00 and [Al/Fe]=0.55, based on the extrema measured in our sample.
As can be easily seen from the plot, this dilution curve fails in reproducing the observed trend. 
More specifically, we can identify three groups of stars: (1) stars that show primordial Li and Al abundances (FG stars); (2) stars with primordial Li but enhancement in their Al content, at different levels; (3) stars with an extreme composition, characterised by paucity of Li and increased Al abundances
(SG stars, with an extreme pattern). 
The majority of the GC stars, that belong to group 2, cannot be explained by diluting the primordial population with the extreme SG, 
because they still exhibit a quite large Li abundance (the curve is indeed a lower envelope to their distribution).
This implies that in order to reproduce their Li abundance, we have to call for a Li production within the stellar polluters.

Alternatively, one possible solution requires the presence of an unobserved population, 
typyfied by [Al/Fe] of approximately $\approx$1.0 dex (the dashed curve in Figure~\ref{fig:M5LiAl}). 
Both this survey and \cite{carretta09b} have failed to identify potential candidates. 
There might be two reasons why this population is unseen:
\begin{itemize}

\item There are no stars formed from the pure ejecta (i.e., with a pollution fraction of 100\%). The processed material coming from 
IM-AGB stars would in this case have Al enhancements of more than [Al/Fe]$\approx$1 dex (and is naturally Na/N/He rich also). This material 
is required to mix with primordial material (and hence become diluted) before the formation of the SG started.
In a recent paper, \cite{dantona12} examined dynamical models where it is possible, in principle, to accumulate 
and mix the ejecta for a time t$_f$ before starting star formation (see Table 1 of their paper).
However, there is no obvious explanation as to the cause of the delay in the star formation events. 
Such a scenario requires that the gas from the AGB stars is collected at the GC centre and remains in a quiescent condition for $\approx$ 40-50 Myr. Star formation is inhibited until the cleared pristine material (swept out from the SN~{\sc ii} explosions) can fall back and mix to produce the   
 the subsequent stellar generation. 

\item These very peculiar stars, characterised by extreme Al over-abundances, should also possess a huge amount of He. 
At a given age, stars with larger amount of He are less massive than their counterpart with normal He (i.e., Y$\sim$0.24).
Considering the metallicity of M5, stars with $Y > 0.35$ will have M$\lesssim$0.5 M$_\odot$ (see \citealt{gratton10a}; \citeyear{gratton10b}) and they might not reach the RGB $tip$. 
In fact, \cite{castellani93} have shown that 
if the mass is smaller than that required to activate the He-core flash, the stars 
will become {\it RGB-manqu{\'e}}. In this circumstance, the He-flash can occur at 
high effective temperatures after stars have left the RGB (the so-called ``hot flashers'') and they eventually move to the {\it blue hook} of the horizontal branch (HB, see e.g., \citealt{moehler12}). 
Unfortunately, as widely discussed in \cite{gratton13}, we can not determine the chemical composition for stars warmer than the {\it Grundahl u-jump} (\citealt{grundahl99}), because of severe sedimentation and radiative levitation effects.

\end{itemize} 

Nevertheless, if this is the case, then an explanation for the two (or even four) Li-poor stars in Figure \ref{fig:M5LiAl} would still be required, perhaps calling for a rare event of extra mixing that begins well before the LF bump luminosity is reached.\footnote{It is also possible that these stars are binaries; we might have
captured post blue straggler stars (BSS) in our RGB sample, which could explain the
Li depletion (\citealt{ryan02}). As already mentioned, the radial velocity is consistent with the cluster mean, but multi-epoch observations to detect possible variations are currently not in hand. We note, however, that 
the fact that all of them are Al-rich seems to suggest that the Li deficiency is somehow related to the multiple population scenario.} 
At the moment there is no compelling evidence to reject them, only because no satisfactory explanation for the cluster's dilution history has presented itself. 
{\it In summary, NGC 5904 displays a degree of complexity that can not be accounted for by a simple dilution model. This cluster demonstrates the presence of three different stellar populations}. This is reminiscent of what \cite{carretta12} discovered in the GC NGC~6752, 
where the intermediate SG stars cannot be explained by simply considering a mixture of primordial composition and (extreme) highly-processed material (i.e., the E stars).


\subsection{The Li spreads in GCs} \label{sec:spreads}

Our investigation into the Li abundance and its potential spread within M12 and M5 has revealed that these GCs behave differently. In M12 we recover a chemical pattern very similar to that previously observed in M4 by DM10 (and corroborated by 
other authors, see e.g., \citealt{mucciarelli11}; \citealt{villanova11}). 
Because FG and SG stars share exactly the same Li abundance, while showing depletion 
in O of more than 50\%, we require that the GC polluters have contributed ashes enriched in Li. As a consequence,  
FRMS (and massive binaries) cannot be responsible for this trend, because the current theory suggests that they carry Li-free ejecta.
On the other hand, in M5 we disclosed the presence of a rather peculiar and complex chemical composition: a simple dilution model fails in reproducing 
the three populations currently co-existing in the cluster. Furthermore, we also need to invoke Li production 
within the polluters to explain the abundance pattern in the majority of the GC stars (i.e., the SG stars which are Al-rich but still Li-rich). 
Crucially, we detected the presence of an extreme population, which is characterised by Al overabundances and Li deficiency
(as expected in the case of hot H burning). 
These stars were not revealed in our M12 sample; however statistics might have played a role in this respect: given that we have 54 stars for which 
Li and Al have been measured, assuming that the fraction of extreme stars is as in M5 (i.e., about 3\%) we would expect to find at least one of those
stars. The probability that we missed all of them is 19\%, which is not negligible. 
This should be regarded as a point of caution in the following discussion.

\begin{figure}
\centering 
 \includegraphics[width=8cm]{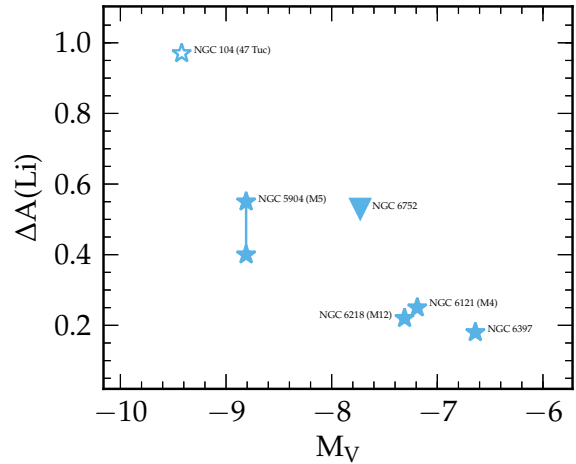}
\caption{$\Delta$A(Li) (peak-to-peak variation) as a function of $M_{\rm{V}}$. For further details about the computation of these values see text.}\label{fig:deltali}
\end{figure}

These two GCs share a very similar metallicity, but they significantly differ in current mass. M5 is much more massive than M12 (and M4).
In order to determine what role GC mass plays in the Li distribution among the different stellar populations, we plot in
Figure \ref{fig:deltali} the internal spread in Li, $\Delta$A(Li), as a function of the absolute visual magnitude (a proxy for the current cluster mass).
We used, as a proxy for the Li distribution, the peak-to-peak variation within each GC. 
This obviously includes the contribution related to the observational uncertainties, implying that the intrinsic Li spread could be smaller than the peak-to-peak values (and virtually zero in cases like M4/12). To evaluate the extent of the
Li spreads in our target GCs, we considered only the stars with magnitudes
fainter than the RGB bump luminosity. The peak-to-peak variation in the Li
abundances is 0.22 dex and 0.55 dex for NGC 6218 and NGC 5904, respectively. Had
we eliminated the four upper limits in M5 the peak-to-peak variation would
be 0.40 dex. As there was a non-negligible probability that we missed corresponding extreme stars in M12,
both values for the Li spread in M5 are plotted in Figure 14. Regardless of which value we adopt, the observed Li spread in M5 is larger than in M12 (recall our observational uncertainties are the same for the two clusters).
We included data for the GC NGC~6397 by selecting a subsample of stars analysed by \cite{lind09}: in order to be as homogeneous as possible with our target giants, we restricted our attention to those RGB stars within approximately 1 magnitude fainter than the bump. This choice, although limiting the sample size, allows us to minimise the impact of the star-to-star difference in the atmospheric parameters (especially in temperatures, which can increase the internal scatter) and guarantees a reliable comparison with our GC giants. We determined a Li spread of 0.18 dex. We proceeded in the same fashion by including Li abundances for a subsample of giants published by \cite{mucciarelli11}. We find the Li variation to be 0.25 dex. 
Unfortunately for the GCs 47 Tuc and NGC 6752, Li abundances in the giant stars have not been determined and we are forced to exploit dwarfs. This should be a point of caution when considering the general trend shown in Figure~\ref{fig:deltali}. Note, in the same context, that the spectra for NGC6752 by \cite{shen10} are characterised by very low S/N ratios, in same cases below 15, possibly suggesting that the spread quoted is over-estimated because of the observational uncertainties.

We detect, for the first time to our knowledge, the existence of an unambiguous 
correlation between the Li variation and the total cluster luminosity 
(i.e., mass; the Pearson's correlation coefficient r $=-0.93$ is significant at more than 99.9\%): {\it the more massive the GC, the larger the Li spread}\footnote{Note that the correlation is still significant at more than 95\% level if we adopt for M5 the value of $\Delta$~A(Li)=0.40, that results from ignoring the four upper limits for the extreme population}. 
This finding seems to suggest that Li production is less efficient in the more massive GCs than in small GCs, 
because any kind of Li replenishment tends to erase the presence of Li-O and/or Li-Na anticorrelation. 
We find that in less massive systems the FG and SG stars are very similar as far as the 
Li content is concerned (even indistinguishable in some cases like M4 and M12). 
We can speculate that in less massive GCs, the polluter mass range might be biased towards the lower end of the IM-AGB stars 
(M$\lesssim$ 6M$_\odot$) whilst in the most massive GCs we expect the upper envelope of the mass distribution to extend beyond 
(M$\gtrsim$ 7 M$_\odot$). 
This is required to account for the considerable p-capture element variations observed in these massive GCs, such as e.g., NGC~2808, where high levels of O depletion and Na enhancement ([O/Fe] down to $\approx -1.00$ and [Na/Fe] up to $\approx +1.00$), as well as 
significant Mg depletion and Al enrichment have been reported.
Regarding Li, any comparison between observed chemical abundances and AGB models must  be done bearing in mind all the uncertainties involved. Li production is indeed extremely sensitive to the 
input physics in the stellar models: in the IM-AGBs there is a very brief phase of Li enrichment at the stellar surface during the first few thermal pulses, with a peak of A(Li)$\sim$4 dex, but the final Li yield depends on when the star loses most of its mass (\citealt{vd10}; \citealt{dorazi13}).
The rate and details of the mass loss are among the most uncertain (and difficult to model) factors in theoretical stellar astrophysics. 

Interestingly, we have demonstrated that the FG and SG stars display the same Li abundances in GCs like M4 and M12. These results imply that the internal polluters must have produced Li in roughly the same amount as the primordial abundances. 
A 5~M$_{\odot}$ AGB stellar model published in
\cite{dorazi13} results in  A(Li)=2.00, by adopting an increased $\alpha_{\rm MLT}$=2.2 and Bloecker (\citeyear{bloecker95}) mass loss law, and A(Li)=2.35 with a ``standard" $\alpha_{\rm MLT}$=1.75 and Bloecker mass loss (see Figure 20 of that paper). Similar Li yields have been found previously by \citealt{vd10}. If we assume that standard Big Bang nucleosynthesis is correct, then FG stars have formed with an initial Li content of A(Li)$\sim$2.7 $-$ 2.8
and they subsequently depleted their Li abundances by a factor of $\approx$3 to the Spite plateau value. 
Unless SG stars somehow deplete Li in a different way compared to FG, they also should have born with A(Li)$\sim$2.7$-$2.8, implying that the polluters must be capable of producing such high Li abundance. Although \cite{dorazi13} found the Li production being lower for AGB models of 5 and 6 M$_{\odot}$, we recall that those yields are highly dependent on mass loss rate. By adopting an even more extreme mass loss than Bloecker, we can reach a higher Li production.
On the other hand, there is still the possibility that standard Big Bang nucleosynthesis is not correct, and that both FG and SG stars formed with Li abundances of A(Li)$\sim$2.1-2.3 dex, with no need to invoke Li depletion in both stellar generations. 
Were this the case, then theoretical models for AGB stars would result in fair agreement with observational measurements without further adjustments to the input physics.
Although there is much evidence to support the predictions of standard Big-Bang Nucleosynthesis  (\citealt{steigman07}), we must give consideration to these possibilities.

The clusters reported in Figure~\ref{fig:deltali} are obviously characterised by a range in metallicity.
The sample includes the metal-poor GC NGC 6397 ([Fe/H] $\approx-$2.0 dex), NGC~6752 ([Fe/H] $\approx-$1.5 dex), the intermediate-metallicity GCs M12, M5 and M4 ([Fe/H] $\approx -$1.3 dex) and one of the most metal-rich GCs in 47 Tuc ([Fe/H] $\approx-$ 0.7 dex). Given the limited sample size available, we cannot robustly infer the level of contribution provided by metallicity on the spread in Li in GCs. We note that two  
GCs with almost the same mass (M4 and NGC 6752), but slightly dissimilar [Fe/H], do actually exhibit Li variations at a different extent.
Irrespective of possible metallicity-related effects, the trend appearing in Figure~\ref{fig:deltali} points to the GC mass as the driving parameter of the anticorrelation. This is evident when we compare GCs with similar metallicity but different masses (e.g., M5 $vs$ M12 and M4). 
It is noteworthy, in this context, that these GCs also display different HB morphology, M4 and M5 being the classical example of the so-called
``second-parameter'' pair (see e.g., \citealt{gratton13} and references therein).

The metal-rich GC 47 Tuc (NGC 104, labelled with a different symbol in Figure~\ref{fig:deltali}) deserves further discussion. 
In fact, in this GC the large scatter detected in the Li abundance seems to be completely unrelated to variations in p-capture elements 
and, more generally unrelated to the presence of multiple stellar populations. 
As discussed in \citet{dorazi10b}, while SG (O-poor) stars are characterised by a low Li abundance, there is a huge spread within the first stellar generation itself,  A(Li) ranging from $\sim$1.5 to $\sim$2.5 (we ignore for the current purposes the presence of a very Li-rich star 
with A(Li)=2.78).
They concluded that this primordial scatter is probably related to the high metallicity of this GC and is likely the Population~{\sc ii} analogue of what is observed in Population~{\sc i} stars with similar atmospheric parameters (such as e.g., in the open cluster M67, \citealt{randich00}).
In addition, 47 Tuc is known to display peculiar behaviour also in terms of light-element variations and their relationship with the cluster 
properties, namely its Na-O anticorrelation 
is relatively short despite the large GC mass. 
It stands as an outlier in the diagrams of IQR\footnote{Interquartile range}[Na/O] $vs$ M$_V$ and $\log$$T_{{\rm eff},{\rm max}}^{{\rm HB}}$ (the maximum temperature on the HB), as derived by \cite{carretta10}. 
By discarding this GC, we still can detect the hint for an anticorrelation but the small statistics prevents us from conclusively confirming its existence. The situation will become clear once a larger sample of GCs (encompassing a wide range in current mass) becomes available in the present survey. 

\subsubsection{The Origin of the Li Spread}

The aim of the work presented here has been to constrain the nature of the progenitor stars that are most responsible for the abundance patterns observed in GCs. The determination of Li abundances in GC stars provide a means to differentiate between those progenitors that are predicted to produce Li and those that are predicted only to destroy it. GCs such as M4 and M12 show very little (if any) difference in the lithium abundance between the two populations, requiring significant Li production in the progenitors (to account for the high Li content still present in SG stars). 
M5 also seems to require some lithium production across the different stellar generations, but not to the same extent. Furthermore, we have identified a possible correlation between the current mass of the GC and the degree of Li spread. We have proposed that this Li variation is related to the amount of Li produced in the progenitors, however there is another contribution we must consider. 

Standard stellar models are unable to produce the observed dispersion in Li unless there is some form of enrichment.
In some instances, particularly when the dispersion is small, the effects of rotation and the instabilities associated with angular momentum loss may be responsible for the dispersion and might in principle affect our ability to put tight constraints on the mass range of the polluting progenitors. 
Models by  \cite{pinsonneault92} and discussed in \cite{deliyannis93} have demonstrated that Li depletion during the pre MS and MS is dependent on the stellar structure (mass, composition, age) and rotational properties (initial angular momentum, timescale for decay).  Models calculated with rotation successfully reproduce the large Li dispersion observed on the MS of open clusters and the small dispersion observed in old metal-poor halo stars.  
Their Population II models, with a representative distribution in mass, metallicity and angular momentum found in the field significantly deplete Li within the first Gyr of the stellar lifetimes. Models that deplete lithium by up to 1 dex produce a dispersion of 0.1 dex and those that deplete Li by 0.3-0.5 dex result in a dispersion of 0.03-0.05 dex.

In the context of the multiple populations in GCs, such a small Li dispersion (at about $\sim$0.1 dex level) may be entirely due to the distribution in angular momentum of second generation stars. It would imply all FG stars formed from a similar angular momentum distribution as those in the halo. Whilst SG stars in more massive clusters form with a larger range of initial angular momenta leading to a larger range in lithium depletion. 
Thus, although Li production within the polluters seems to be mostly responsible for the total internal spread in Li abundances, a further (minor) contribution due to e.g., MS depletion cannot be discarded. In fact, observational data for M4 and M12 (e.g., \citealt{mucciarelli11} or this study) are in agreement with a very limited Li variation (within 0.1 dex), which might be explained in terms of phenomena unrelated to the multiple populations frameworks, such as e.g., MS depletion due to rotational-induced mixing.

\section{Summary and concluding remarks}\label{sec:summary}

Lithium provides rare insight into not only the internal processes of stars but also the internal chemical enrichment of GCs.
The abundance patterns within these old stellar aggregates are most straightforwardly explained by the presence of multiple populations, whereby a first generation of stars has polluted the medium from which a second generation form. Within this second generation it is still possible for distinct chemical populations to form (see \citealt{carretta09a} and their PIE definitions). Because of its fragility and thus the special conditions required for its production, Li may serve as a 
unique tracer of the nature of stars that provided the intra-cluster enrichment.   
Current stellar theory predicts that it is possible for intermediate-mass asymptotic giant branch stars to produce lithium via the Cameron-Fowler mechanism, whereas fast-rotating massive stars and massive binaries will produce Li-free ejecta. How lithium correlates with other p-capture species will reveal whether this element has been produced between the different stellar generations and thus help to identify the progenitors. 
Following on from the work of DM10 who focussed on NGC 6121, we have presented lithium and [Al/Fe] abundances in stars on the RGB of the GCs NGC 6218 and NGC 5904. Our findings can be summarised as follows:

In the GC NGC 6218 (M12), any Li variation is less than observational errors and is consistent with no Li variation between the two populations. Thus, whilst the cluster displays clear [Na/Fe] (over 1 dex) and [O/Fe] (over 1.2 dex) variations, stars across their respective populations remain indistinguishable according to their Li abundance.  
These variations in Na and O are accompanied by variations in the [Al/Fe] abundance.  The (anti)correlations that form between p-capture nuclei are expected when hydrogen burning at high temperatures has been in operation (T$\gtrsim$20 MK). 
Because the large variations in p-capture elements are not accompanied by corresponding changes
in Li, Li production must have
occurred across the different stellar generations. With our current understanding of stellar evolution, such a result favours a major contribution from IM-AGB progenitors, as also found by DM10 in the analogous GC NGC 6121 (M4).

In the GC NGC 5904 (M5), we are unable to statistically confirm Li variation with O or Na; however there is a hint for a  Li-Al anticorrelation. We anticipate that the small number of stars with both Li and Na (or O) measured is hindering our ability to detect a relationship between these species. 
There are possibly four (two confirmed) stars that are very lithium poor for their evolutionary phase. They may be due to non-standard evolution or perhaps members of an extreme (third) population in this cluster, that displays Li deficiency 
and Al enhancement. 
Given the presence of a Li-Al anticorrelation in NGC 5904, we have fit a dilution model as per \citet{2006A&A...458..135P} to explain the chemical history of the cluster. When we mix the composition representative of the primordial population with that of four candidate extreme members, we are unable to account for the abundances of a majority of the 
cluster stars (which are Li-rich and Al-rich). 
To do so would require the pristine material to be combined with an unobserved population that has [Al/Fe] of approximately $\approx$ 1.0 dex. 
If this is the case, by trying to reproduce a majority of the cluster members we still require an explanation for the (four) extreme population candidates. Thus, NGC~5904 exhibits a level of complexity much higher than its more standard siblings, like M4 and M12. This result is not surprising when we consider the HB morphology of this GC.

There is a clear anticorrelation between the internal Li spread and the current GC mass. 
Although \cite{pinsonneault90} and \cite{pinsonneault92} have shown that rotation during the MS can produce dispersions in the Li abundance, we attribute the spread to production in the progenitors.
Li production appears less efficient in the more massive GCs than in smaller GCs, pointing to a different mass range of the stellar polluters involved in the internal enrichment. 
Further surveys will help constrain this indication on more robust grounds and reveal what role metallicity plays in the lithium spread found within GCs.

\acknowledgments{This work made extensive use of the SIMBAD, Vizier, and NASA ADS databases. 
We thank Simon Campbell, Maria Lugaro, and Enrico Vesperini for very helpful discussions. VD is an ARC Super Science Fellow. 
EC, RGG, SL, and YM acknowledge partial support by PRIN INAF 2011 ``Multiple populations in globular 
clusters: their role in the Galaxy assembly'' (PI E. Carretta).
We thank the referee for a very careful reading of the manuscript and for her/his very helpful comments and suggestions}

\bibliographystyle{apj}

\end{document}